\documentclass[prd,onecolumn,nopacs,nokeys]{revtex4}
\usepackage{amsfonts}
\usepackage{amsmath}
\usepackage{amssymb}
\usepackage{bbm}
\usepackage[utf8]{inputenc}
\usepackage{tensor}
\usepackage{slashed}
\usepackage[centertableaux ]{ytableau}
\usepackage{hyperref}
\usepackage{natbib}
\usepackage{enumerate}
\usepackage{braket,mleftright}
\usepackage{graphicx}
\usepackage{subfigure}
\usepackage{cleveref}

\begin{document}

\title{Production of dark matter in association with a Higgs boson via exclusive photon fusion in $pp$ collisions at $\sqrt{s}=13$~TeV.}

\author{M. A. Arroyo-Ureña}
\email{marco.arroyo@fcfm.buap.mx}
\affiliation{Facultad de Ciencias F\'isico-Matem\'aticas, Benem\'erita Universidad Aut\'onoma de Puebla, C.P. 72570, Puebla, M\'exico,}
\affiliation{Centro Interdisciplinario de Investigación y Enseñanza de la Ciencia, Benemérita Universidad Autónoma de Puebla, Av. San Claudio y Prol. 24 sur, Col. San Manuel Ciudad Universitaria, 72570, Puebla, Puebla, M\'{e}xico. }

\author{H. Hern\'{a}ndez-Arellano} 
\email{col539008@colaborador.buap.mx}
\affiliation{Centro Interdisciplinario de Investigación y Enseñanza de la Ciencia, Benemérita Universidad Autónoma de Puebla, Av. San Claudio y Prol. 24 sur, Col. San Manuel Ciudad Universitaria, 72570, Puebla, Puebla, M\'{e}xico. }

\author{I. Pedraza}
\email{isabel.pedraza@correo.buap.mx}
\affiliation{Centro Interdisciplinario de Investigación y Enseñanza de la Ciencia, Benemérita Universidad Autónoma de Puebla, Av. San Claudio y Prol. 24 sur, Col. San Manuel Ciudad Universitaria, 72570, Puebla, Puebla, M\'{e}xico. }

\author{S. Rosado-Navarro}
\email{sebastian.rosado@protonmail.com}
\affiliation{Centro Interdisciplinario de Investigación y Enseñanza de la Ciencia, Benemérita Universidad Autónoma de Puebla, Av. San Claudio y Prol. 24 sur, Col. San Manuel Ciudad Universitaria, 72570, Puebla, Puebla, M\'{e}xico. }

\author{T. A. Valencia-P\'erez}
\email{tvalencia@fisica.unam.mx}
\affiliation{Instituto de F\'isica\\
Universidad Nacional Aut\'onoma de M\'exico, Ciudad de M\'exico, C.P. 04510, M\'exico.}

\begin{abstract}
In this work we study the production of a dark matter (DM) particle in association with a  Higgs boson via a central exclusive photon-fusion initiated process. We explore this type of production through the Inert Doublet Model plus a complex Singlet (IDMS), where an extension of the  Standard Model by an additional $U(1)_X$ gauge symmetry and a $SU(2)$ inert scalar doublet gives rise to a DM candidate $\chi$. This particular process involves the collision of two protons exchanging two colorless particles (in our case, photons), from which a central process occurs. Such  interaction can be detected in the LHC using forward proton detectors, where the resulting missing mass spectrum can be observed after proton reconstruction, thus allowing a search for physics Beyond the Standard Model (BSM). We present results for different values of the difference of masses of a heavy scalar coming from the complex singlet, the DM candidate and the Higgs boson,  $\Delta = M_S - M_\chi - M_h$, which is the phase space available for the  final state in the central exclusive process.
\end{abstract}

\maketitle

\section{Introduction}

The nature of dark matter (DM) is one of the greatest challenges that concern the fields of cosmology, astrophysics and particle physics today. Currently, evidence of its existence is based only on its coupling through gravity, measured from observations such as galaxy rotational curves and the cosmic microwave background. Throughout the last fifty years, precise observations allow us to collect a number of general properties for DM, involving its contribution to the energy density in the universe as well as its distribution in certain astrophysical regions (for a compelling review see \cite{cirelli2024darkmatter}). In a effort to find new ways to observe the potential interactions of DM with ordinary matter, there are three general paths to take: direct detection, indirect detection and searches at colliders.

DM detection at collider experiments is performed through the observation of a missing energy signal, after DM production from a collision, similar to the way neutrinos are detected. Such an event can be detected only if other visible particles are produced along DM during the collision. An interesting prospect is to observe this signal in Central Exclusive Production (CEP) in proton-proton collisions, in which we have $p p \to p + \psi + p$, where $\psi$ is a defined central state. The outgoing protons remain intact and are scattered in small angles, and all their lost energy goes into the $\psi$ state \cite{Heinemeyer:2008pk,FP420RD:2008jqg,Albrow:2010yb,Heinemeyer:2012hr}. They can then be detected by near-beam detectors such as the CMS-TOTEM Precision Proton Spectrometer (CT-PPS) or the ATLAS Forward Proton spectrometer at the Large Hadron Collider (LHC). The reconstruction of the momentum loss and transverse momentum of the outgoing protons permits, in turn, the reconstruction of the mass of the centrally produced system $\psi$ with great resolution using the missing mass technique, regardless of the mechanism involving its production \cite{Albrow:2000na}.

In this work we look at the process where the central state $\psi$ is produced by photon fusion. In a previous work \cite{Arroyo-Urena:2019zah}, an extension to the Standard Model (SM) with an additional $U(1)_X$ gauge symmetry allows the introduction of a DM candidate through one of the doublets, which is considered to be inert, thus called the Inert Doublet Model (IDMS). In this framework, the scalar field $S_X$ can be produced in photon fusion and consequently decay into a Higgs boson and the DM candidate $\chi$. 

The structure of this paper is as follows. We describe in detail the theoretical framework for this analysis in section \ref{theory}. Section \ref{ParamSpace} is dedicated to discussing the surveyed parameter space of the model, and the viability of this framework as a DM candidate by requiring consistency with relic density and direct detection limits. Then, in section \ref{collider}, we examine the photon-fusion induced CEP process $pp \to p  (\gamma\gamma\to S)  p \to p  (S\to h  \chi)  p$ and calculate the cross section as well as the missing mass spectrum. Finally, we present our conclusions and perspectives on feasibility of observing such a signal in proton-proton colliders with near-beam detectors.

\section{Theoretical framework} \label{theory}
In this study we rely on an extension of the SM, called Inert Doublet Model plus a complex singlet (IDMS), which introduces a $U(1)_X$ gauge symmetry, a $SU(2)$ scalar doublet $\Phi_2$ and a singlet scalar field $S_X$ to the SM gauge group $\mathcal{G}_{\text{SM}}=SU(3)_C\times SU(2)_L \times U(1)_Y$.  
The singlet scalar $S_X$ is responsible for breaking the $U(1)_X$ symmetry into $\mathcal{G}_{\text{SM}}$. We make $\Phi_2$ into an inert scalar doublet in order to propose a dark matter candidate. The scalar doublet $\Phi_2$ has a Vacuum Expectation Value (VEV) equal to zero to try the stability of the DM candidate. However, this requirement is not enough to guarantee stability, thus a mechanism to control its possible decays into other particles is necessary. There are, at least, two options to achieve that: a discrete $Z_2$ symmetry or the $U(1)_X$ gauge symmetry \cite{Haber:2018iwr,Bonilla:2014xba,Krawczyk:2015xhl}, which will be described further in this text. First, we begin with the specifications of the model.
\subsection{Scalar fields}
The scalar fields can be written as follows:
\begin{eqnarray}
\Phi_1 &=& \left( \begin{array}{c}
\phi_1^+\\  \nonumber
\frac{1}{\sqrt{2}}(\upsilon+\phi_1+i\xi_1)
\end{array} \right),\\
\Phi_2 &=& \left( \begin{array}{c}
\phi_2^+\\
\frac{1}{\sqrt{2}}(\phi_2+i\xi_2)
\end{array} \right),\\ \nonumber
S_X &=& \frac{1}{\sqrt{2}}(v_x+s_x+i\xi_x),
\label{eq:scalar_reps}
\end{eqnarray}
where their assignments under the $G_{\text{SM}}\times U(1)_X $ group are given by:
\begin{eqnarray}
\Phi_1&\to&(\textbf{1}, \textbf{2}, 1/2, x_1),\\ \nonumber
\Phi_2&\to&(\textbf{1},\textbf{2},1/2,x_2),\\ \nonumber
S_X&\to&(\textbf{1},\textbf{1},0,x).
\end{eqnarray}
The first (second) entry corresponds to the representation under $SU(3)_C$ ($SU(2)_L$), while the hypercharge (charge) under $U(1)_X$ is written in the third (last) entry.
The spontaneous symmetry breaking (SSB) is achieved through the chain:
\begin{equation*}
G_{\text{SM}}\times U(1)_X  \xrightarrow{\langle S_X \rangle} G_{\text{SM}} \xrightarrow{\langle \Phi_1 \rangle} SU(3)_C\times U(1)_{\text{EM}},
\end{equation*}
where $\langle S_X \rangle = v_x/\sqrt{2}$ and  $\langle \Phi_1 \rangle^T = (0,v/\sqrt{2})$ with $v=246.22$~GeV.
\subsection{Scalar potential}
The most general, renormalizable and gauge invariant potential is given by
\begin{eqnarray}
V &=&\mu _{1}^{2}\Phi _{1}^{\dag }\Phi _{1}+\mu _{2}^{2}\Phi _{2}^{\dag}\Phi _{2}+\mu _{x}^{2}\mathcal{S}_{X}^{\ast }\mathcal{S}_{X}+\left[ \mu _{12}^{2}\Phi_{1}^{\dag }\Phi _{2}+h.c.\right]  \nonumber \\
&+&\lambda _{x}\left( S_{X}^{\ast }S_{X}\right) ^{2}+\lambda _{1}\left( \Phi _{1}^{\dag }\Phi _{1}\right) ^{2}+\lambda_{2}\left( \Phi _{2}^{\dag }\Phi _{2}\right) ^{2} \nonumber \\
&+&\lambda _{3}\left( \Phi_{1}^{\dag }\Phi _{1}\right) \left( \Phi _{2}^{\dag }\Phi _{2}\right)+\lambda _{4}\left\vert \Phi _{1}^{\dag }\Phi _{2}\right\vert ^{2}+ \left[\lambda _{5}\left( \Phi _{1}^{\dag }\Phi _{2}\right) ^{2}\right.\nonumber \\
&+&\left. \lambda _{6}\left(\Phi _{1}^{\dag }\Phi _{1}\right) \left( \Phi _{1}^{\dag }\Phi _{2}\right)
\right.   \left. +\lambda _{7}\left( \Phi _{2}^{\dag }\Phi _{2}\right) \left( \Phi_{1}^{\dag }\Phi _{2}\right) +h.c.\right]   \nonumber \\
&+&\left( S_{X}^{\ast }S_{X}\right) \left[ \lambda _{1x}\left( \Phi_{1}^{\dag }\Phi _{1}\right) +\lambda _{2x}\left( \Phi _{2}^{\dag }\Phi_{2}\right) \right]  \nonumber \\
&+&\left[ \lambda _{12x}\left( \Phi _{1}^{\dag }\Phi _{2}\right) \left( S_{X}^{\ast }S_{X}\right) +h.c.\right],
\label{potential}
\end{eqnarray}
where $\mu_{1,\,2}^2$, $\lambda_{1,...,4,\,1x,\,2x}$ are real parameters and, in general, $\mu_{12}^2$, $\lambda_{5,\,6,\,7,\,12x}$ are complex parameters. Note that $\mu_2^2>0$ because the DM candidate arises from $\Phi_2$, which has $<\Phi_2> = 0$. We note that the terms proportional to $\Phi _{1}^{\dag }\Phi _{2} S_{X}$ or $\Phi _{2}^{\dag }\Phi _{1} S_{X}$ could induce decays of the DM candidate. To solve this problem, we consider that $x_2-x_1\pm x\neq0$ in order to keep these terms non-invariant under gauge symmetry. Thus, the parameters that accompany these terms must be zero to recover the gauge invariance and at the same time eliminate the couplings that are responsible for a decay of DM candidate at two neutral scalars.

By considering the $ \{ \phi_1, \,s_x,\, \phi_2,\, \xi_2 \}$ basis, and after the SSB, the scalar mass matrix can be written as follows
\begin{equation}
\bf{M_{0}^2}=\left(\begin{array}{cccc}\rm M_{11} &\rm M_{12} &\rm M_{13} &0 \\\rm M_{12} & \rm  M_{22} & \rm M_{23} &0\\\rm M_{13} & \rm M_{23} &\rm M_{33} &\rm M_{34} \\0 & 0 &\rm M_{34} &\rm M_{44}\end{array}\right),
\label{mass0}
\end{equation}
where
\begin{eqnarray}
\rm M_{11}&=&2\lambda_1 \upsilon^2,\;\;\;\;\;  \rm M_{12}=\lambda_{1x}\upsilon\, \upsilon_x,  \;\;\;\;\;\rm M_{13}=\frac{1}{2}\lambda_{6}\upsilon^2,\nonumber\\
\rm M_{22}&=&2\lambda_{x}\upsilon_x^2, \;\;\;\;\;\rm M_{23}=\frac{1}{2}\lambda_{12x}\upsilon \upsilon_x,\nonumber\\ 
\rm M_{33}&=&\mu_2^2+\frac{1}{2}(\lambda_3+\lambda_{4}+\textrm{Re}[\lambda_{5}])\upsilon^2+\frac{1}{2}\lambda_{2x}\upsilon_x^2,\nonumber\\
\rm M_{34}&=&-\textrm{Im}[\lambda_{5}]\upsilon^2,\nonumber \\
\rm M_{44}&=&\mu_2^2+\frac{1}{2}(\lambda_3+\lambda_{4}-\textrm{Re}[\lambda_{5}])\upsilon^2+\frac{1}{2}\lambda_{2x}\upsilon_x^2.
\label{massmatrix}
\end{eqnarray}
Once the $\bf{M_{0}}$ matrix is diagonalized and neutral scalars are rotated to physical states, the $\rm M_{13,\,23}$ matrix elements induce a mixing between neutral scalars and DM candidate, which compromises the stability of the DM candidate. This means that the terms proportional to $ \Phi _{1}^{\dag }\Phi _{2}$ must be removed, otherwise the DM candidate will be unstable. For $x_1=x_2$, the terms proportional to $ \Phi _{1}^{\dag }\Phi _{2}$ in the scalar potential are gauge invariant. Then, it is required to introduce an additional discrete $Z_2$ symmetry for the doublets to eliminate these terms in the potential. Moreover, for $x_1\neq x_2$, the gauge invariance of the $U(1)_X$ symmetry guarantees the stability for the DM candidate. In either case, we will assume that $\lambda_6=\lambda_7=0$ in order to maintain the invariance under $Z_2$ or $U(1)_X$ symmetries. It is worth noting that $\lambda_{12x}$ also has implications for the stability of the dark matter candidate. Nevertheless, the kinematic condition $\Delta = M_S - M_{\chi} - M_h > 0$, which enforces $M_S > M_{\chi}$, ensures that stability is preserved.
\subsubsection{$Z_2$ symmetry and $x_1=x_2$ case}
From Eq. \ref{potential}, we note that the terms proportional to $\Phi_{1}^{\dag }\Phi _{2}$ respect invariance under $U(1)_X$. However, it is necessary to invoke a $Z_2$ discrete symmetry to remove them, i.e.,  $\Phi_1\rightarrow \Phi_1$ and $\Phi_2\rightarrow -\Phi_2$. By considering the last assignment for the doublet,  $\rm M_{13}=\rm M_{23}=0$. Then, the mass matrix for the neutral scalar, Eq. (\ref{mass0}), is diagonalized by
\begin{equation}
\left(\begin{array}{c} h \\ S\end{array}\right)=\left(\begin{array}{cc}\cos\alpha_1 & -\sin\alpha_1 \\ \sin\alpha_1 & \cos\alpha_1
\end{array}\right) \left(\begin{array}{c} \phi_1 \\ s_x \end{array}\right)
\label{alpha1}
\end{equation}
and
\begin{equation}
\left(\begin{array}{c} \chi \\ A\end{array}\right)=\left(\begin{array}{cc}\cos\alpha_2 & -\sin\alpha_2 \\ \sin\alpha_2 & \cos\alpha_2
\end{array}\right) \left(\begin{array}{c} \phi_2 \\ \xi_2 \end{array}\right),
\end{equation}
where $\tan\alpha_{1,2}=\frac{\rho_{1,2}}{1+\sqrt{1+ \rho_{1,2}^2}}$ with $\rho_1=\frac{\lambda_{1x}\upsilon \upsilon_x}{\lambda_1\upsilon^2-\lambda_{x}\upsilon_x^2}$ and $\rho_2=\frac{-\textrm{Im}[\lambda_5]}{\textrm{Re}[\lambda_5]}$ \cite{Cabral-Rosetti:2017mai}. Therefore, the scalar masses are given by
\begin{eqnarray}\label{HiggsBosonMass}
M_{S,h}^2&=&\lambda_1\upsilon^2+\lambda_{x}\upsilon_x^2\pm(\lambda_1\upsilon^2+\lambda_{x}\upsilon_x^2)\sqrt{1+\rho_1^2},\\ \nonumber
M_{H^\pm}^2&=&\mu_2^2+\frac{1}{2}(\lambda_3 \upsilon^2+\lambda_{2x}\upsilon_x^2)\label{eqmCHmass},\\ \nonumber
M_{A,\chi}^2&=&M_{H^{\pm}}^2+\left(\frac{\lambda_4}{2}\pm|\lambda_5|\right) \upsilon^2.
\end{eqnarray}
At this point, we propose to $\chi$ as our DM candidate. 
A fact to be mentioned is that the model can induce the $\chi\to H^{\pm}W^{\mp}$ decay, whose $\chi H^{\pm}W^{\mp}$ coupling is shown in table \ref{FR-IDMS}. In this case, we demand that the masses must satisfy $M_{H^\pm}^2\,(M_{A}^2)>M_{\chi}^2$, which is achieved, from Eq. \ref{HiggsBosonMass}, via the following constraint: 
\begin{equation}
    \upsilon_x^2>\frac{2(M_h^2+\mu_2^2)}{4\lambda_x-\lambda_{2x}}.
\end{equation}
If one considers the particular case Im$[\lambda_5]=0$, the only possible effects of CP-violation come from $\lambda_{12x}$, and therefore $\tan\alpha_2=0$, which has as a consequence that $\cos\alpha_2=1$. This condition results in no mixing between the neutral parts of the second scalar field  and the complex singlet with the physical states.
\subsubsection{$U(1)_X$ gauge symmetry, $x_1\neq x_2$ case}
The condition $x_1\neq x_2$ also allows the stability of the DM candidate. In addition to the constraints $\lambda_6=\lambda_7=0$, in this case, $\lambda_5$ must be also zero.

\subsection{Yukawa Lagrangian}
The most general Yukawa Lagrangian can be written as follows
\begin{equation}
\mathcal{L}_{Yukawa}=\sum_{i,j=1}^{3}\sum_{a=1}^{2}\left( \overline{Q}
_{Li}^{0}Y_{aij}^{0u}\widetilde{\Phi }_{a}u_{Rj}^{0}+\overline{Q}
_{Li}^{0}Y_{aij}^{0d}\Phi _{a}d_{Rj}^{0} \right.
+\left.\overline{\ell}_{Li}^{0}Y_{aij}^{0\ell}
\Phi _{a}\ell_{Rj}^{0}+h.c.\right),  \label{yukawa}
\end{equation}

where $Y_{a}^{0f}$ are the $3\times 3$ Yukawa matrices, with $f=u,d,\ell$; $Q_{L}$ and $\ell_{L}$ denote the left-handed fermion doublets under $SU(2)_L$ and $f_{R}$ represents to the right-handed fermion singlets. The zero superscript in the fermion fields stands for the interaction basis. The DM stability is lost if the couplings $Y_{2ij}^{0f}$ appear in the Eq. (\ref{yukawa}). These Yukawa couplings can be eliminated by the correct assignment of values for charges under the $Z_2$ and $U(1)_X$ symmetries, as previously done.

When one considers a discrete $Z_2$ symmetry (and $x_1=x_2$), we have $Y_{2ij}^{0f}=0$ to respect such a $Z_2$ symmetry. However, the couplings $Y_{1ij}^{0f}\neq 0$ if the assignment of the $U(1)_X$ charges for the fermions satisfy
\begin{equation}
\mp x_1-x_q+x_{u,d}=0
\label{xquark}
\end{equation}
and
\begin{equation}
x_1-x_l+x_{e}=0,
\label{xlepton}
\end{equation}
where $x_{q,l}$ are the $U(1)_X$ charges of left-handed doublet fermions, meanwhile, $x_{u,d,e}$ are the $U(1)_X$ charges of right-handed fermions.

When $x_1\neq x_2$, we set the $U(1)_X$ charges as $\mp x_2-x_q+x_{u,d}\neq0$ and  $x_2-x_l+x_{e}\neq0$ in order to remove the couplings $Y_{2ij}^{0f}$ in Eq. \ref{yukawa}. Note that $\Phi_1$ also satisfies Eqs. \eqref{xquark} and \ref{xlepton} to give mass to the fermions as in the case of the SM. The relevant Feynman rules of IDMS are shown in Table \ref{FR-IDMS}, where we have defined $\rho\lambda_{12x}\equiv g_{Sh\chi}$, with $\rho=v_x\cos(2\alpha_1)\approx v_x$. A key outcome of our analysis, which will be elaborated upon in the following sections, is that $v_x$ naturally resides at the TeV scale. This phenomenological requirement has profound implications for the structure of the scalar potential, as it necessarily forces $\lambda_{12x}$ to acquire small values in order to maintain the perturbativity and naturalness of the coupling $g_{Sh\chi}$, ensuring that it remains below unity.
\begin{table}[!htb]
	\caption{IDMS couplings involved in the calculations of this work.
		We define $\lambda_{345}=\lambda_{3}+\lambda_{4}+2\lambda_{5}$. For $Z^{\prime}f_i\bar{f}_i$ coupling we consider the limit when the kinetic mixing term $\varepsilon\to0$.\label{FR-IDMS}}
	\centering{}%
	\begin{tabular}{c c}
		\hline
		Coupling & Expression\tabularnewline
		\hline
		\hline
		$hf_{i}\bar{f}_{i}$ & $\frac{m_{f_{i}}}{\upsilon}\cos\alpha_{1}$\tabularnewline
		\hline
		$hH^{-}H^{+}$ &$(\lambda_3\cos\alpha_{1}+\lambda_{2x}/2)\upsilon$    \tabularnewline
		\hline
		$hW_{\mu}^{-}W_{\nu}^{+}$ & $gm_{W}\cos\alpha_{1}g_{\mu\nu}$\tabularnewline
		\hline
		$h\chi\chi$ & $(\lambda_{345}\cos\alpha_{1}+\lambda_{2x}/2)\upsilon$\tabularnewline
		\hline
		$Z^{\prime}_{\mu}f_{i}\bar{f}_{i}$ & $\frac{g_x}{2}\left(1-\gamma^{5}\right)\gamma^{\mu}$\tabularnewline
		\hline
		$Z_{\mu}^{\prime}\chi\chi$ & $\frac{g_x}{2}(p_{Z_{\mu}^{\prime}}-p_{\chi})^{\mu}$\tabularnewline
		\hline
		$Sf_{i}\bar{f}_{i}$ & $\frac{m_{f_{i}}}{\upsilon}\sin\alpha_{1}$\tabularnewline
		\hline
		$SW^-_{\mu}W^+_{\nu}$ & $gm_W\sin\alpha_{1}g_{\mu\nu}$\tabularnewline
		\hline
		$S\chi\chi$ & $\lambda_{2x}\upsilon_{x}\cos\alpha_{1}-\lambda_{345}\upsilon\sin\alpha_1$\tabularnewline
		\hline
		$\chi (A) H^{\pm}W_{\mu}^{\mp}$ & $i \frac{g}{\sqrt{2}}(p_{H^{\pm}}-p_{\chi(A)})^{\mu}\cos\alpha_2$\tabularnewline
		\hline
        $\chi S h$ & $\rho\lambda_{12x}\equiv g_{Sh\chi}$\tabularnewline
  		\hline
	\end{tabular}
\end{table}

In order to guarantee an anomaly free model it is remarkable to highlight that the fermion charges under the $U(1)_X$ symmetry must satisfy the triangle anomaly equations \cite{Mantilla:2016lui}. The anomaly cancellation requirements for fermion charges $x_{f}$, for $f=q,u,d,l,e$, are shown in table \ref{table_fermion_x} as a function of $x_q$ \cite{Arroyo-Urena:2019zah}.
\begin{table}
	\caption{Relations between fermions charges under $U(1)_X$ to guarantee the anomaly cancellation.}
	\label{table_fermion_x}
	\centering{}%
	\begin{tabular}{c c}
		\hline
		Field & $U(1)_X$\tabularnewline
		\hline
		\hline
		$q_L$ &  $x_q$\tabularnewline
		\hline	
		$u_R$ & $x_u=4x_q$ \tabularnewline
		\hline
		$d_R$ &  $x_d=-2x_q$ \tabularnewline
		\hline
		$l_L$ &  $x_l=-3x_q$ \tabularnewline
		\hline
		$e_R$ &  $x_e=-6x_q$ \tabularnewline
		\hline
	\end{tabular}
\end{table}

\subsection{Gauge bosons interactions}
The kinetic term for the $U(1)_{Y,\,X}$ gauge symmetries can be written as follows:
\begin{equation}
\mathcal{L}_{\rm Kin} = - \frac{1}{4} \hat{B}_{\mu\nu} \hat{B}^{\mu\nu}- \frac{1}{4}\hat{Z}_{0\mu\nu}^{\prime}\hat{Z}^{\prime 0\mu\nu} + \frac{1}{2}\frac{\varepsilon}{\cos\theta_{W}} \hat{B}^{\mu\nu} \hat{Z}_{0\mu\nu}^{\prime},
\label{eq:kin}
\end{equation}
where, $ \hat{B}^{\mu\nu}$ and  $\hat{Z'_0}^{\mu\nu}$ are the field strength tensors defined by $\hat{F}_{\mu\nu}=\partial_\mu \hat{F}_\nu-\partial _\nu \hat{F}_\mu$ for $\hat{F}_{\nu}=\hat{B}_{\nu},\,\hat{Z}_{0\nu}$ \cite{Lee:2013fda,Davoudiasl:2012ag}. Note the fact that the mixing term between $\hat{B}_{\mu\nu}$ and $\hat{Z}_{0\mu\nu}^{\prime}$ is allowed by the gauge invariance. However, this mixing term can be eliminated by the following redefinition

\begin{eqnarray}
 \left( \begin{array}{c}
 Z'_{0\mu}\\
B_\mu
\end{array} \right)
=
 \left( \begin{array}{cc}
\sqrt{1-\varepsilon^2/\cos^2\theta_W} & 0\\
-\varepsilon/ \cos^2\theta_W&1 \end{array} \right) \left( \begin{array}{c}
 \hat{Z'}_{0\mu}\\
\hat{B}_\mu
\end{array} \right).
\end{eqnarray}
The hatted fields contain the kinetic mixing term and, according with experimental measurements, $\varepsilon$ is smaller than $\mathcal{O}(10^{-3})$ \cite{Abel:2008ai}. After SSB, the gauge bosons, in the mass basis, are given by

\begin{eqnarray}
Z_{0\mu}&=&\hat{Z}_{0\mu}+\varepsilon\tan\theta_W \hat{Z}_{0\mu}^{\prime},\\ \nonumber
A_\mu&=&\hat{A}_\mu-\varepsilon \hat{Z}_{0\mu}^{\prime},\\ \nonumber
Z_{0\mu}^\prime&=&\hat{Z}_{0\mu}^{\prime}.
\end{eqnarray}

Meanwhile, the interaction gauge-scalar fields reads
\begin{equation}
\mathcal{L}_\text{scalar} = \sum_{a=1}^2 | D_\mu \Phi_a |^2 + | D_\mu S_X |^2,
\end{equation}
where $D_{\mu}$ is the covariant derivative given by
\begin{eqnarray}
D_{\mu} =  \left(  \partial_\mu + i g^{\prime} Y \hat B_\mu + i g T_3 \hat W_{3 \mu} + i g_{x} Q^{\prime}_i \hat Z^{\prime}_{0\mu} \right) ,
\end{eqnarray}
with $g_X$ and $Q^{\prime}_i$ denoting the coupling constant and the charge under the $U(1)_X$ symmetry, respectively. After the SSB, we obtain both the mass and mixing terms:
\begin{equation}
\mathcal{L}_\text{scalar}=\frac{1}{2}M_{Z^\prime}^2Z^{\prime0} Z^{\prime0}+\frac{1}{2}M_{Z}^2Z^0Z^0-\Delta^2Z^{0} Z^{\prime0}+...,
\end{equation}
where
\begin{eqnarray}
M_{Z^\prime}^2 &=&\left( \frac{g^\prime\varepsilon}{2\cos\theta_W}+g_x x_1 \right)^2 \upsilon^2+g_x^2x^2\upsilon_x^2,\\ 
M_{Z}^2 &=& g^2\frac{\upsilon^2}{4\cos^2\theta_W},
\end{eqnarray} \label{eq:mZ'-mZ}
and
\begin{equation}
\delta^2=\frac{1}{2}g_Z\left(\frac{g^\prime\varepsilon}{2 \cos\theta_W}+g_x x  \right)\upsilon^2.
\end{equation}
In order to cancel the mixing term, the following rotation is required
\begin{eqnarray}
\left(  \begin{array}{c}
Z \\ Z^\prime
\end{array} \right)
=
\left( \begin{array}{cc}
\cos\xi & -\sin\xi \\ \sin\xi & \cos\xi
\end{array} \right)
\left(
\begin{array}{c}  Z^0 \\ Z^{\prime0}
\end{array} \right),
\end{eqnarray}
where the mixing angle $\xi$ satisfies the expression $\tan 2\xi = \frac{2\delta^2}{m^2_{Z^0}-m^2_{Z^{'0}}}$, and has been constrained to the interval $|\xi|<10^{-3}$ \cite{Bouchiat:2004sp}.

\section{IDMS parameter space} \label{ParamSpace}

\begin{figure}
    \centering
    \includegraphics[width=0.5\linewidth]{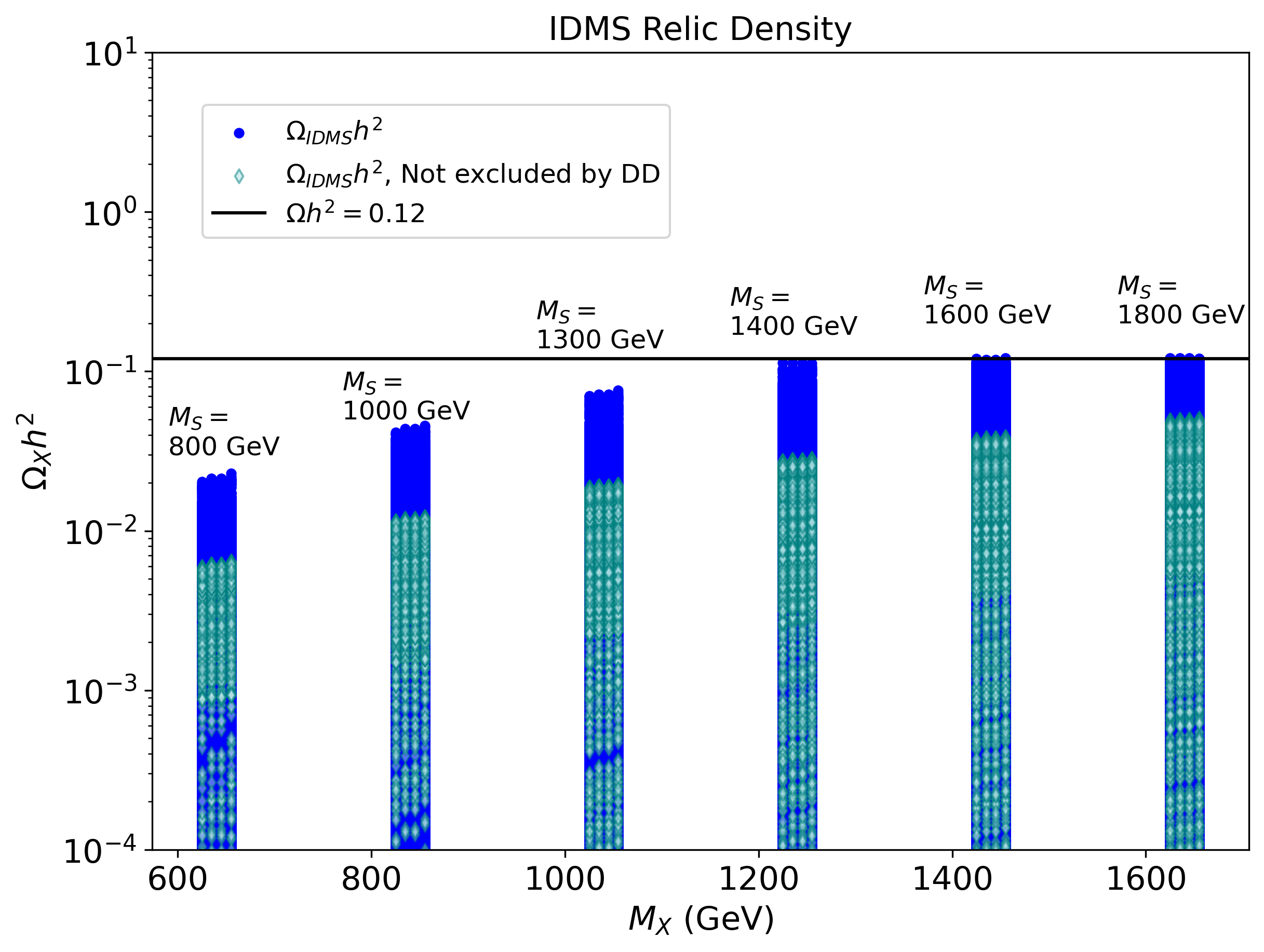}
    \caption{The value of the relic density ($\Omega_{IDMS} h^{2}$) in relation to the value of the dark matter candidate mass, $M_{\chi}$. The solid circle markers represent the all the different parameter configurations studied, while the diamond markers represent those not excluded by direct detection limits (\cite{lux-zeplin}). }
    \label{fig:RelicD}
\end{figure}

Firstly, we constrain the parameter space by surveying a wide range of values, restricting the value of the dark matter candidate relic density to the most recent estimation, $\Omega h^2 = 0.1200 \pm 0.0012$ \cite{Planck:2018vyg}. For this purpose we make use of the MicrOMEGAs code \cite{Alguero:2023zol} to calculate $\Omega h^2$ for $M_{S} = 800, 1000, 1200, 1400, 1600, 1800$ GeV, each for $\Delta = 20, 30, 40 , 50$ GeV. Within the same framework, we calculate the dark matter - nucleon cross-section and compare with the current most restringent limits for this mass range set by the LUX-ZEPLIN experiment \cite{lux-zeplin}. In Fig. ~\ref{fig:RelicD} we present a plot of the relic density for each value of the dark matter candidate mass $M_{\chi}$. Complying with both the PLANCK relic density estimation and the LUX-ZEPLIN direct detection limits, for the mass range chosen, we conclude that taking our proposition as a dark matter candidate, an additional source must be added to account for the full relic density. 

\begin{figure}
    \centering
    \includegraphics[width=0.5\linewidth]{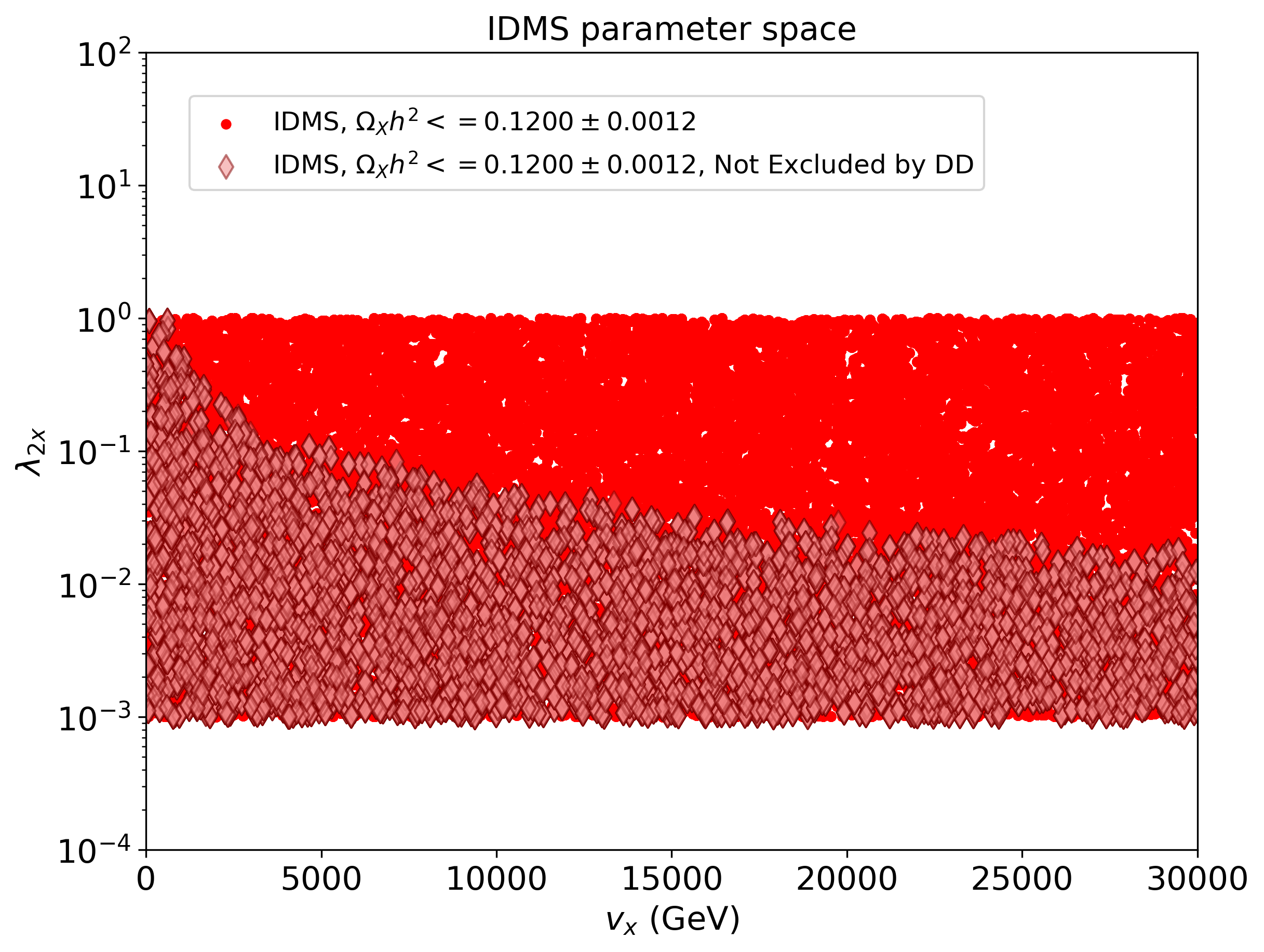}
    \caption{Values of $\lambda_{2x}$ and $v_{x}$ requiring relic density to be below or equal to the PLANCK estimation (solid circles), and with the additional direct detection bound consistency requirement (diamonds).} 
    \label{fig:vvs_vs_lambda2x}
\end{figure}

Some of the parameters surveyed (i.e. $v_{x},\lambda_1, \lambda_2$) are not impacted by requiring consistency with PLANCK and LUX-ZEPLIN's limits, while for $\lambda_{2x}$ particularly, we see a restriction which depends with the value of $v_x$, seen in Fig. ~\ref{fig:vvs_vs_lambda2x}. Notably, the direct detection constraints impose a reduction on the allowed values of $\lambda_{2x}$, particularly for larger $v_x\sim30$ TeV, where $\lambda_{2x}$ is restricted to $\mathcal{O}(10^{-2})$. This behavior is closely related to the IDMS couplings presented in Table~\ref{FR-IDMS}. The DM-nucleon scattering cross section, receives contributions from Higgs-mediated diagrams (in the $t-$channel) involving the $h\chi\chi$ coupling, as well as from diagrams mediated by the heavy scalar $S$ via the $S\chi\chi$ interaction. Both of these couplings, shown explicitly in Table~\ref{FR-IDMS}, depend on $\lambda_{2x}$: the $S\chi\chi$ coupling is proportional to $\lambda_{2x}v$, while the $h\chi\chi$ coupling contains a term proportional to $\lambda_{2x}$ as well.

The range of the parameters surveyed and the impact of demanding consistency with direct detection constraints is presented in Table ~\ref{tab:parameter-ranges}.

\begin{table}[!htb]
    \caption{Range of parameters surveyed, and the impact of requiring consistency with PLANCK and LUX-ZEPLIN limits. $\lambda_3,\lambda_4$ and $\lambda_5$ are uniformely reduced in range, while $\lambda_{2x}$ restricted region depends on the parameter $v_x$. The latter relation can be seen in Fig. ~\ref{fig:vvs_vs_lambda2x}.}
    \label{tab:parameter-ranges}
    \centering
    \begin{tabular}{c c c}
    \hline
    \textbf{Parameter} & \textbf{\begin{tabular}[c]{@{}c@{}}Surveyed \\ range\end{tabular}} & \textbf{\begin{tabular}[c]{@{}c@{}}Consistent with\\ Direct Detection\end{tabular}} \\ \hline
    \hline
    $v_x$                 & $[0,30]$ TeV                                                       & $[0,30]$ TeV                                                                                      \\ \hline
    $\lambda_1$,$\lambda_2$              & $[10^{-3},1]$                                                      & $[10^{-3},1]$                                                                                      \\ \hline
    $\lambda_3$,$\lambda_4$,$\lambda_5$           & $[10^{-3},1]$                                                      & $[10^{-3},10^{-1}]$                                                                    \\ \hline
    $\lambda_{2x}$                & $[10^{-3},1]$                                                      & \begin{tabular}[c]{@{}c@{}}$[10^{-3}, \sim10^{-2}]$ \\ for $v_x = 30$ TeV\end{tabular}                                                       \\ \hline
    $\lambda_{12x}$               & $[10^{-7},1]$                                                      & $[10^{-7},1]$                                                                                      \\ \hline
    \end{tabular}
\end{table}

\begin{figure}
    \centering
    \includegraphics[width=0.5\linewidth]{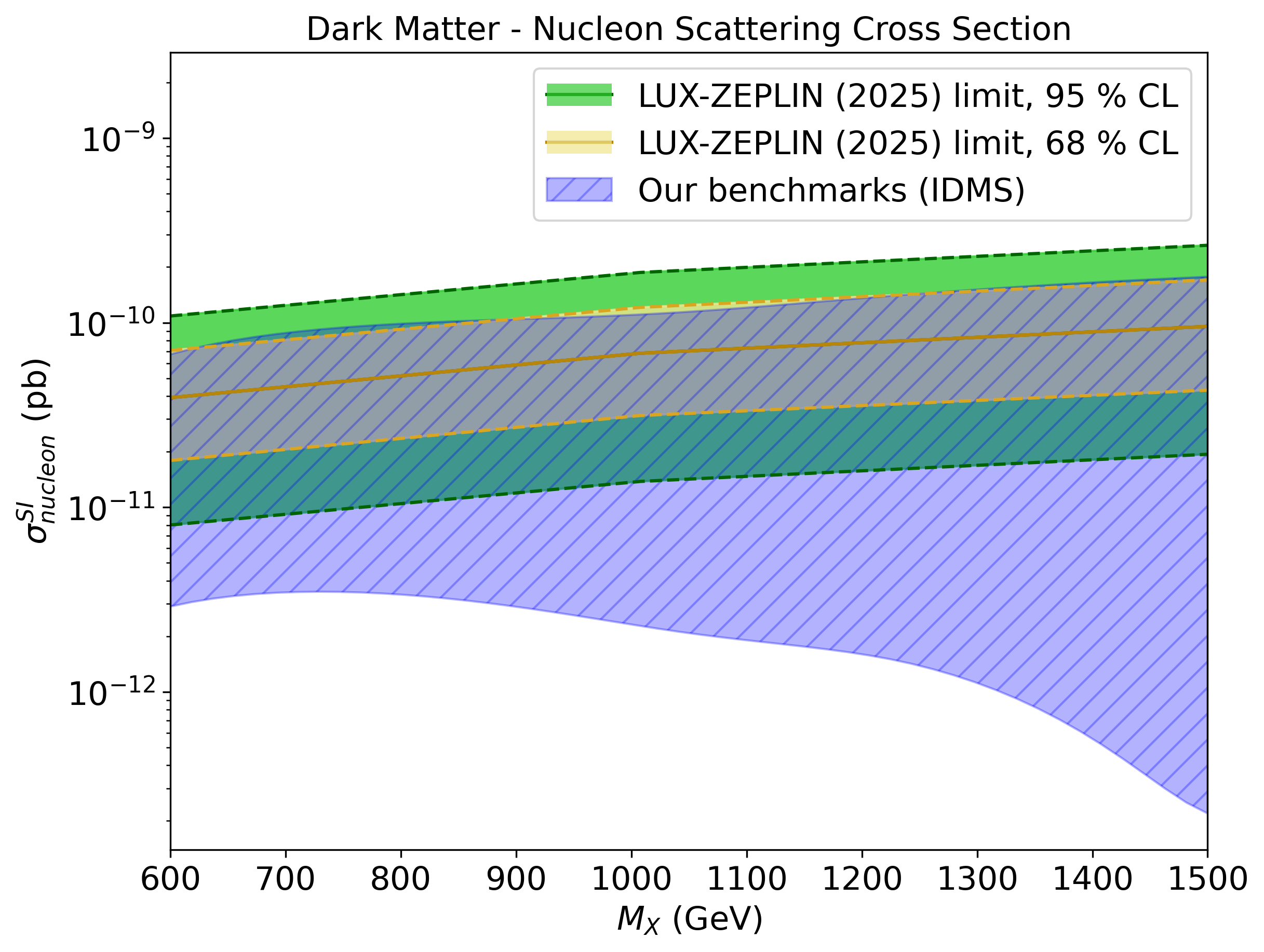}
    \caption{Dark Matter - Nucleon scattering cross section for the chosen parameter space in comparison with the LUX-ZEPLIN limits (\cite{lux-zeplin}).}
    \label{fig:directdetection}
\end{figure}

In order to test the model against collider data constraints, we chose a set of benchmark parameter configurations, requiring each of them to comply with the relic density and direct detection limits. These benchmark points were utilized to perform the simulations described in the next section. For comparison purposes, the spin-independent IDMS-nucleon cross-section is plotted in Fig.~\ref{fig:directdetection}, along side the upper limit (at 68\% and 95\% C.L.) derived from \cite{lux-zeplin}. This result is intimately connected to the $h\chi\chi$, $S\chi\chi$ interactions (see Table \ref{FR-IDMS}), as the spin-independent scattering cross section receives dominant contributions from Higgs-mediated diagrams, as well as from heavy scalar $S$.

\section{Collider analysis} \label{collider}
\begin{figure}[!htb]
	\centering
	\includegraphics[width=0.3\textwidth]{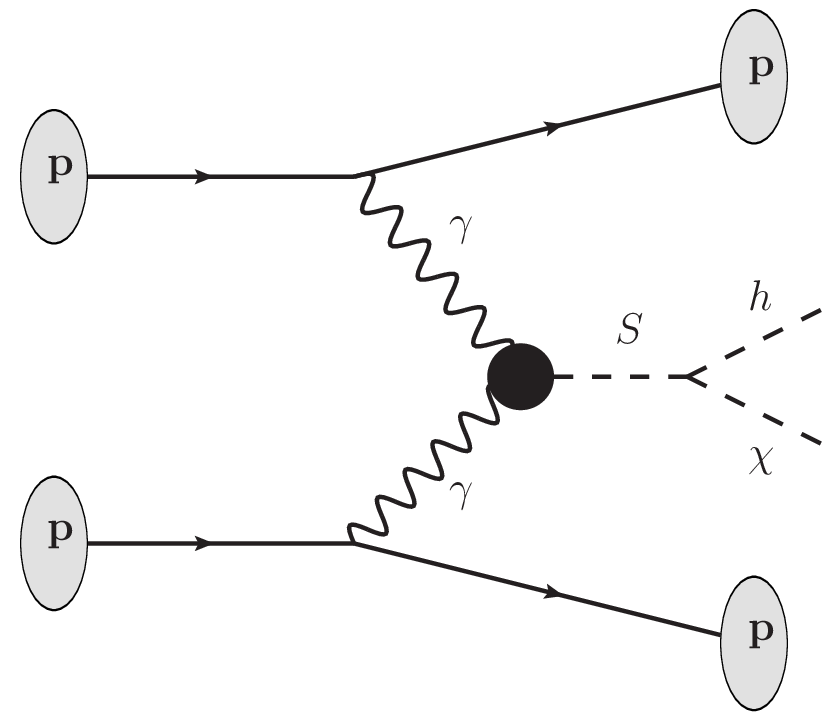}
	\caption{Feynman diagram of the signal $pp\to p(S\to h\chi)p$.} 	\label{FeynmanDiagramSGN}
\end{figure}

We are interested in analyzing the process $pp \to p(S \to h \chi)p$, where the incoming protons scatter via photon exchange, inducing an electromagnetic interaction that produces a scalar particle $S$. This scalar subsequently decays into a Higgs boson $h$ and a DM candidate $\chi$. The corresponding Feynman diagram for this process is shown in Fig.~\ref{FeynmanDiagramSGN}. The final state is characterised by two very forward scattered protons, the decay products of the Higgs boson and missing transverse energy from the undetected DM particle.

A search for DM particles produced in association with a Higgs boson decaying to a bottom quark-antiquark pair in proton-proton collisions at $\sqrt{s}=13$ TeV was reccently reported by CMS collaboration in Ref.~\cite{CMS:2026gje},
corresponding to an integrated luminosity of 101 fb$^{-1}.$ In particular, the analysis is focused in two channels: $(a)$ $gg\to A\to ah,$ with $\,a\to\chi\chi$ and $h\to b\bar{b}$, where $A$ and $a$ are a heavy and a light pseudoscalar particle, respectively, whereas $\chi$ denotes a DM candidate; and $(b)$ $qq\to Z^{\prime}h,$ with $Z^{\prime}\to\chi\chi$ and $h\to b\bar{b}.$ In our work we consider an alternative production mechanism via exclusive photon fusion, which provides a clean experimental signature due to the presence of forward protons.

\subsection{Signal cross sections} \label{subsec:crosssections}

\begin{figure}[!htb]
\centering
\subfigure[]{\includegraphics[scale=0.4]{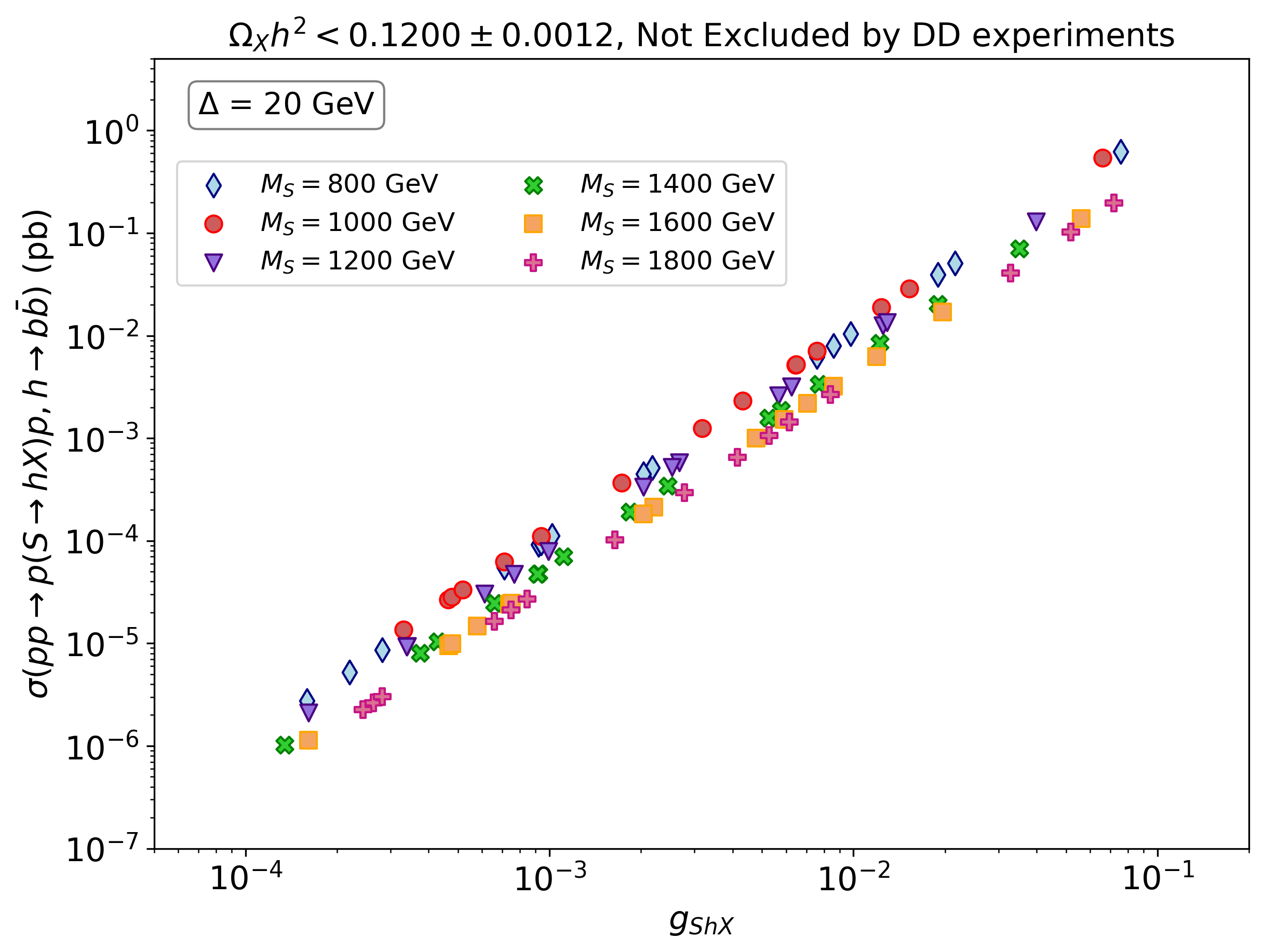}}
\subfigure[]{\includegraphics[scale=0.4]{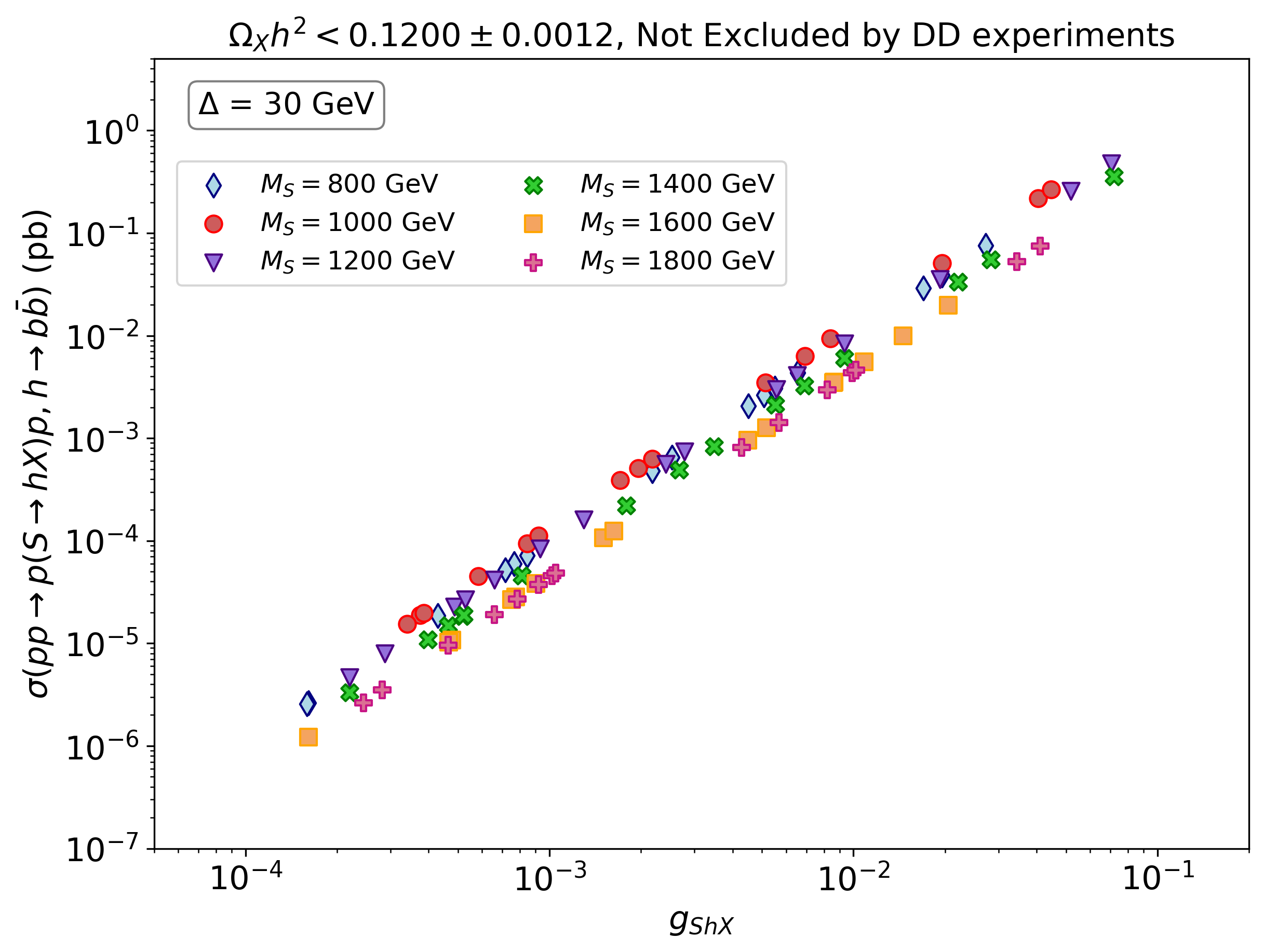}}
\subfigure[]{\includegraphics[scale=0.4]{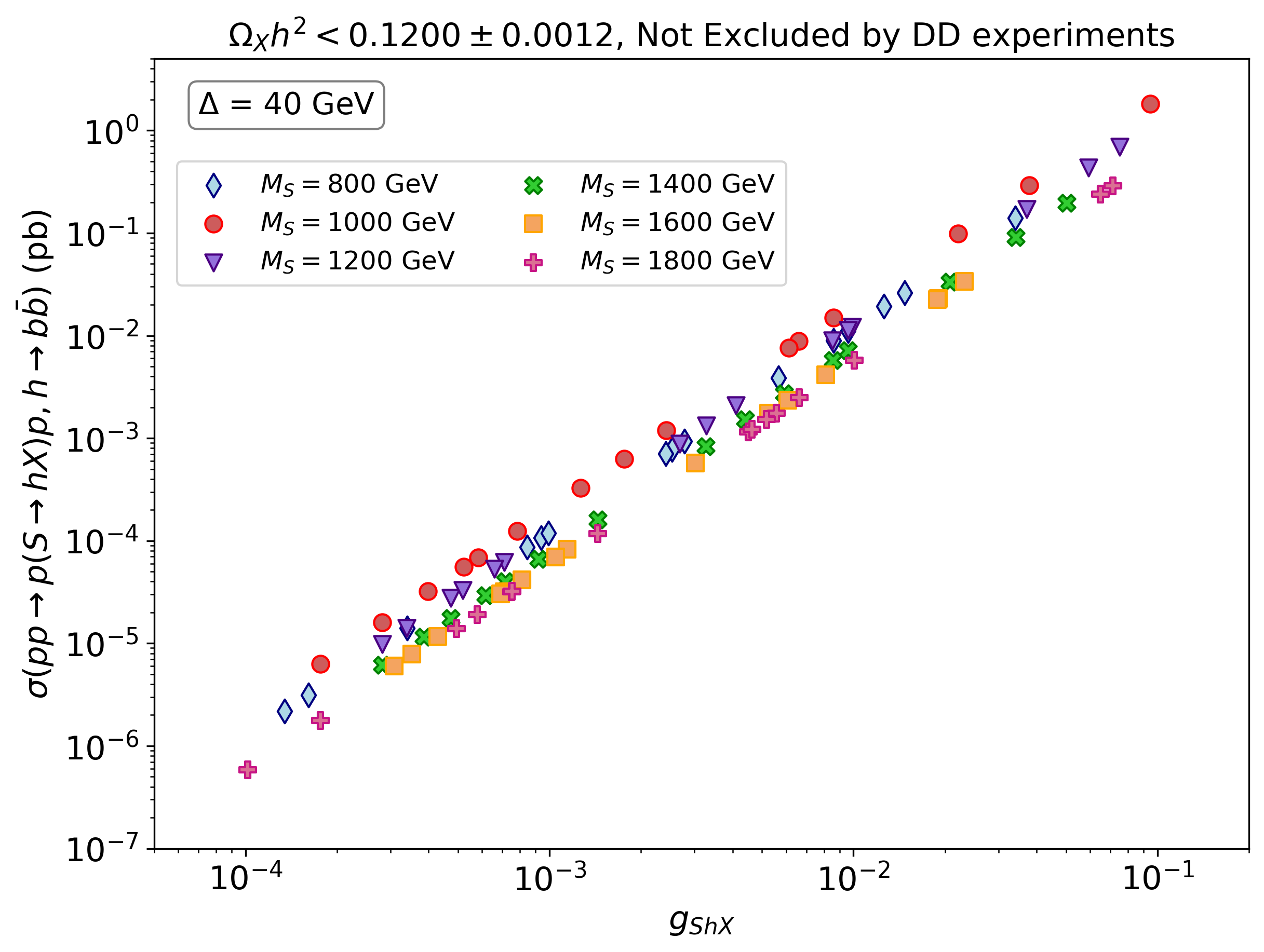}}
\subfigure[]{\includegraphics[scale=0.4]{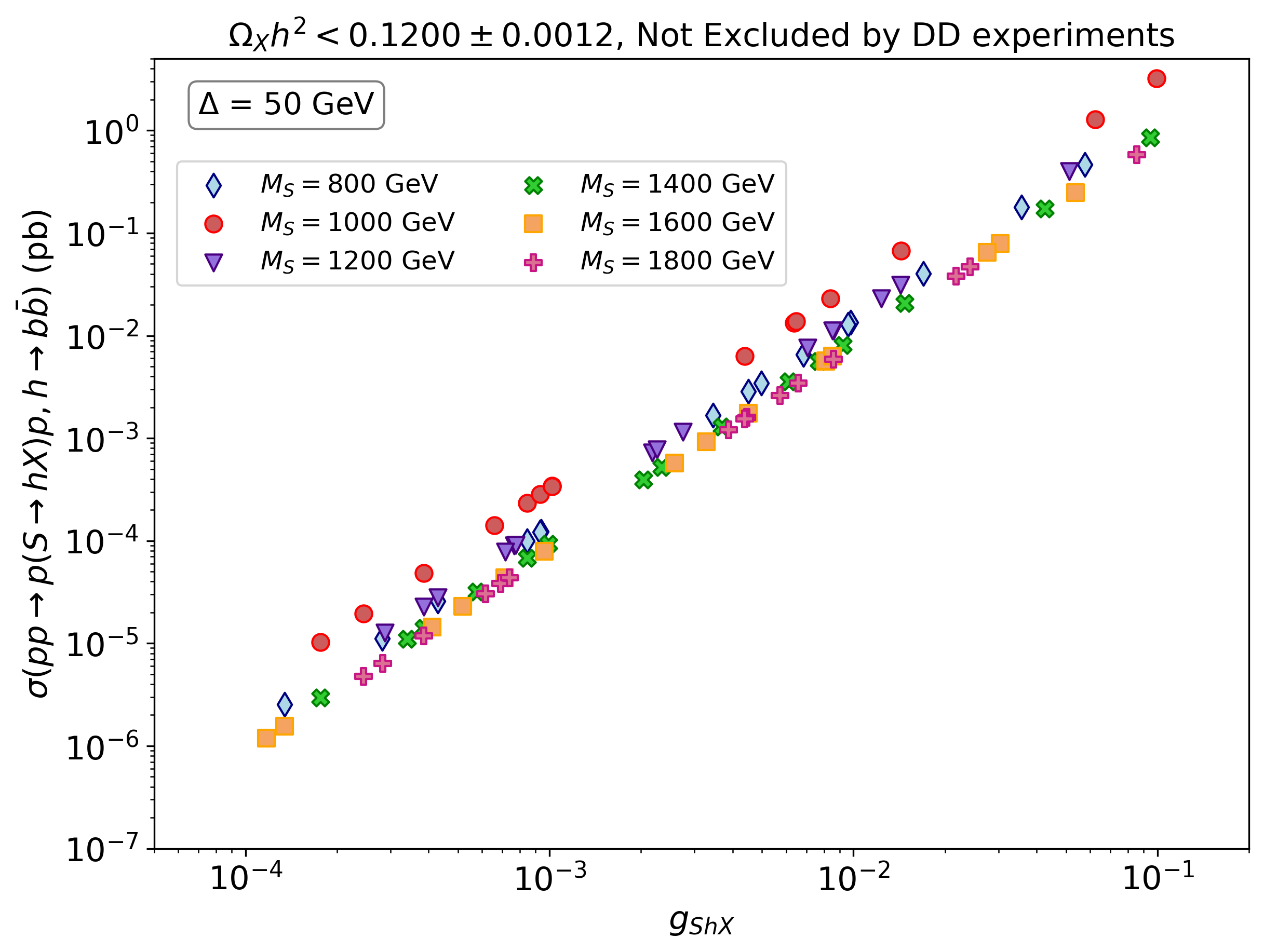}}
\caption{Cross section $\sigma(pp\to p(S\to h\chi)p)$ as a function of the coupling constant $g_Sh\chi$ for $M_S=800,\,1000,\,1200,\,1400,\,1600,\,1800$ and (a) $\Delta=20$ GeV,  (b) $\Delta=30$ GeV, (c) $\Delta=40$ GeV, (d) $\Delta=50$ GeV. 
}
\label{XSs}
\end{figure}

Figure~\ref{XSs} shows the production cross section of the signal as a function of the coupling $g_{Sh\chi}$ for scalar masses $M_S=800,\,1000,\,1200,\,1400,\,1600,\,1800$ GeV and mass splitting values  $\Delta=20,\,30,\,40,\,50$ GeV, while the mass of the DM candidate is determined by the relation $\Delta = M_S - M_\chi - M_h$. It is of utmost importance to highlight that this relationship prohibits the decay of the DM candidate since it is fulfilled that $M_S>M_{\chi}$. As expected from kinematic considerations, the cross section exhibits a monotonically increasing behavior with both the coupling $g_{Sh\chi}$ and the available phase space $\Delta$. Conversely, for a fixed value of $g_{Sh\chi}$ and $\Delta$, the cross section decreases as $M_S$ increases, which is a direct consequence of the phase-space suppression for heavier scalar masses. The colored regions in each panel highlight the parameter space points that simultaneously satisfy the observed relic abundance from the PLANCK satellite and the most recent spin-independent direct detection exclusion limits from the LUX-ZEPLIN experiment, thus representing the most phenomenologically viable regions of our model.


 Insofar as our computation scheme is concerned, we used $\texttt{LanHEP}$~\cite{Semenov:2014rea} to implement IDMS and generate the corresponding \texttt{UFO} files~\cite{Degrande:2011ua}, and subsequently we compute the cross section using $\texttt{MadGraph5}$ \cite{Alwall:2014hca}. 

\subsection{Comparison with current experimental limits} \label{subsec:limits}

\begin{figure}
    \centering
    \includegraphics[width=0.5\linewidth]{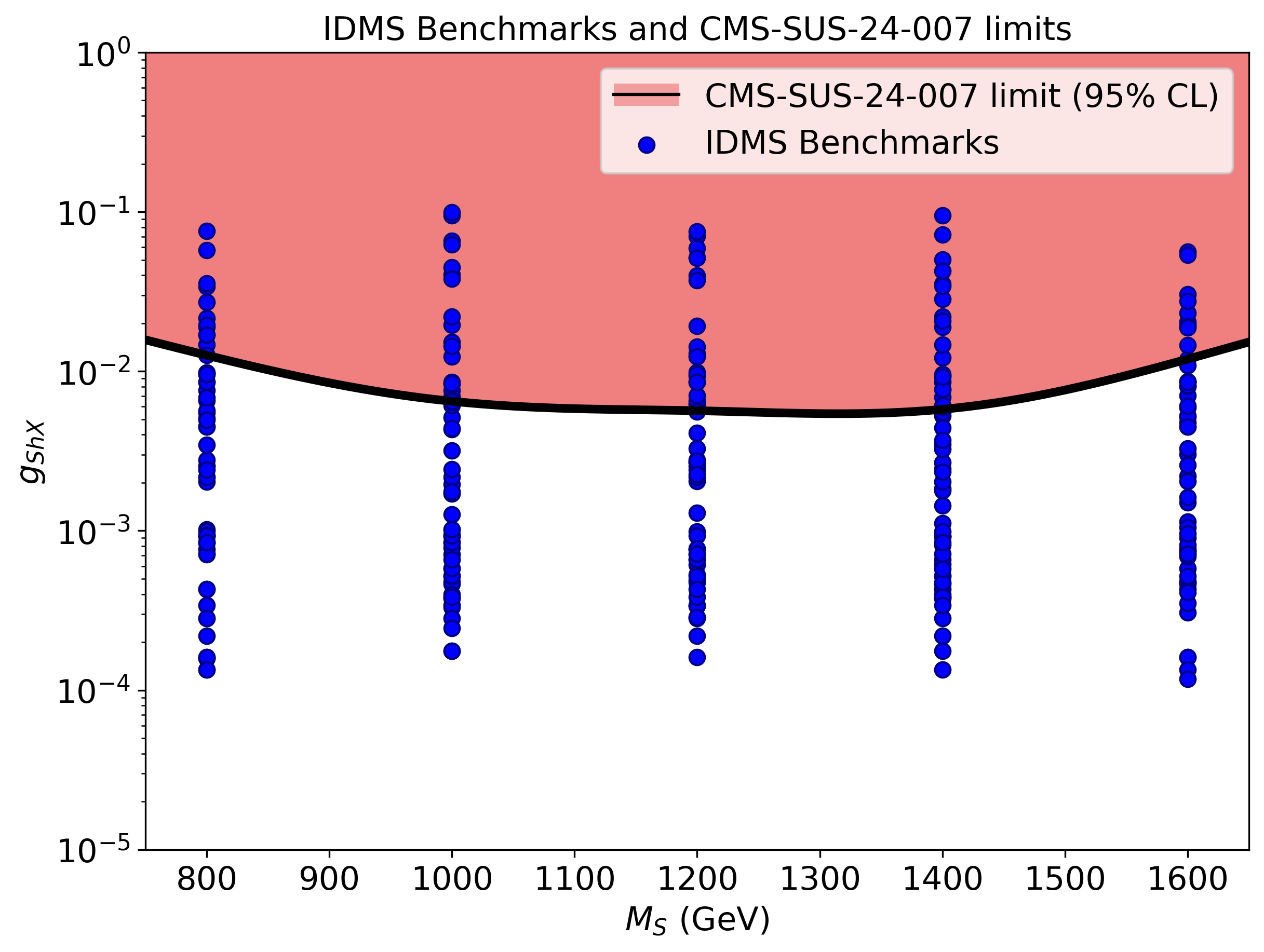}
    \caption{Benchmark values of $g_{Sh\chi}$ and $M_{S}$ compared to the limit (95\% C.L.) extracted from Figure 9b of \cite{CMS:2026gje}. The benchmark points inside the shaded region are excluded by the CMS analysis limits, while those outside remain allowed and could be probed in future searches with higher luminosity or improved analysis techniques. } 
    \label{fig:sus24007-limits}
\end{figure}

To assess the viability of our parameter space, we compare our benchmark points with the exclusion limits derived from the CMS search aforementioned \cite{CMS:2026gje}. In Fig.~\ref{fig:sus24007-limits} these limits are represented with a solid black line, while also showing the values of the coupling $g_{Sh\chi}$ and the scalar mass $M_S$ for the points that survive the relic density and direct detection constraints. In this scenario, the constraints are complied with if $g_{Sh\chi} \sim 10^{-3}$.

\section{Conclusions}
In this work we have explored the production of a dark matter candidate ($\chi$) in association with a Higgs boson via central exclusive photon fusion in proton-proton collisions at $\sqrt{s} = 13$ TeV, within the framework of the Inert Doublet Model extended by a complex singlet (IDMS). This scenario provides a viable dark matter candidate arising from an inert scalar doublet, whose stability is ensured either by a discrete $Z_2$ symmetry or by the $U(1)_X$ gauge symmetry, which depend on the charge assignment $x_i$ of the scalar fields. A key aspect of our analysis is the kinematic condition $\Delta = M_S - M_{\chi} - M_h > 0$, which guarantees that the decay $S \to h \chi$ is kinematically allowed while simultaneously preserving the stability of the dark matter particle. We have shown that the parameter $\lambda_{12x}$, although potentially dangerous for DM decay, does not compromise stability under this condition.
\\
Using a scan over the IDMS parameter space, we identified benchmark points consistent with the observed relic density from PLANCK and the most recent spin-independent direct detection limits from the LUX-ZEPLIN experiment. These points were used to compute the signal cross section for the exclusive process $pp \to p (S \to h\chi) p$ as a function of the coupling $g_{Sh\chi}$, the scalar mass $M_S$, and the mass splitting $\Delta$. Our results indicate that the cross section increases with both $g_{Sh\chi}$ and $\Delta$, while decreasing for larger $M_S$ due to phase-space suppression.
\\
When comparing with existing collider constraints, particularly the CMS search for dark matter produced in association with a Higgs boson decaying to $b\bar{b}$, we find that a significant portion of our benchmark parameter space remains viable. The exclusive photon fusion mechanism offers a complementary and particularly clean experimental signature due to the presence of forward protons, which can be detected by near-beam spectrometers such as the CT-PPS or the ATLAS Forward Proton detector.
\\
Thus, the IDMS provides a well-motivated framework for dark matter, and the exclusive production channel studied here represents a promising avenue for probing this model at the LHC. Future runs with higher luminosity and improved forward detector performance could further constrain or potentially discover a signal in the missing mass spectrum. A more detailed detector-level simulation, including background estimation and realistic acceptance effects, would be valuable to assess the full discovery potential of this channel.
\section*{Acknowledgments}
	The work of M. A. Arroyo-Ure\~na, H. Hern\'andez-Arellano and T. Valencia-P\'erez is supported by ``Estancias posdoctorales por M\'exico (SECIHTI)" and ``Sistema Nacional de Investigadoras e Investigadores" (SNII). T.V.P. also acknowledges support from the UNAM project PAPIIT IN111224 and the CONAHCYT project CBF2023-2024-548. S. Rosado-Navarro thanks to Vicerrector\'ia de Investigaci\'on y Estudios de Posgrado through ``Centro Interdisciplinario de Investigaci\'on y ense\~nanza de la Ciencia".

\bibliography{refs1}

\end{document}